# X-ray Bright Sources in the Field of Active Galactic Nuclei MKN 205

Badie Korany[1,2]

Email: badiekorany@yahoo.com, baewiss@uqu.edu.sa

[1]Department of Physics, Faculty of Applied Science, Umm Al-Qura University, Saudi Arabia
[2]Department of Astronomy, National Research Institute of Astronomy and Geophysics (NRIAG), 11421 Helwan, Cairo, Egypt

**Abstract:** A three bright X-ray non-target sources are detected in the field of the seyfert 1 galaxy MKN 205. These sources are classified as optically Early-type galaxy, BLAGN, and NELG (NED, SIMBAD, and AXIS). The spectrum analysis is made for these objects using thermal models and non-thermal models, modified by interstellar absorption. In some objects we can not distinguish between the thermal and non-thermal models of the hard components from the spectrum alone. The presence of intrinsic absorption is tested by photoelectric absorption at the redshift of the sources, and we assumed the flux distribution is affected by the intrinsic absorption in some sources. A black body component used to test the presence of soft excess in some spectra, which has been interpreted as primary emission from the accretion disc or as secondary radiation form the reprocessing of the hard X-ray in the surface layers of the disc.

1- Introduction

Mkn 205 is a nearby ($z= 0.071$) low luminosity radio quiet quasar with an intriguing Fe Kα emission line complex. X-ray emission is an important tool for the investigation of the gravitational evolution of the cosmos. Most of the sources making up the cosmological X-ray background turn out to be different types of Active Galactic Nuclei, AGNs, where the X-ray emission is due to the accretion of matter onto a supermassive Black Hole, the remainder being due to radiation from the hot gas in the deep potential wells of galaxy clusters (Korany et. al. 2009). Modeling of the X-ray spectrum of the background radiation in terms of these individual sources requires a mixture of objects, displaying different amounts of low energy absorption in their spectra. Highly absorbed objects will thus emit the bulk of their X-ray radiation at energies above 2 keV where the amount of available spectral data is limited. The high sensitivity at energies up to 10 keV, provided by XMM-Newton satellite, will offer the possibility to perform a comprehensive study of the soft and hard X-ray spectra of samples of serendipitous X-ray sources in deep extragalactic fields. This will allow the investigation of the spectral properties of objects which due to their low flux or hard spectrum could not be observed by previous X-ray instruments. In this work we study some non-target X-ray sources in MKN 205which is a nearby (z= 0.071) low luminosity radio quiet quasar with an intriguing Fe Kα emission line complex (Sibasish Laha, et. al. 2019)
field observed by XMM-Newton.

In the present paper, we report on the detection and spectral analysis of X-ray observation of the field of MKN205, taken by XMM-Newton observations. We organized the paper as follows: The X-Ray observations and data reduction are presented in section 2, section 3 is devoted to the spectral analysis, while the results are summarized and concluded in section 4.

## 2- Observation and Data Reduction

The identifications presented in this paper correspond to X-ray sources found in XMM-Newton data of MKN 205. The observations with the EPIC MOS (Turner et al.2001) and PN (Struder et al., 2001) detectors were split into 3parts, each of which was exposed for 17 ksec duration, to test a variety of sub-window modes. Three observations each was made with the MOS 1 and 2 cameras, in Full Window, Partial Window 2 and Partial Window 3 modes. For the PN, two observations were made in Full Window mode and one in Large Window mode. The re-processed data were reduced with the SAS software, using EMCHAIN and EPCHAIN ; further filtering was then performed using xmmselect. A light-curves for the observations (PN, MOS1, and MOS2 ) created to check for flaring high background periods, which are best visible above 10 keV. An images extracted in the energy bands 0.2-0.5, 0.5-2.0, 2.0-4.5, 4.5-7.5 and 7.5-12.0 keV with binsize 22 in cases of MOS1 and MOS2, and 82 in PN case. The edetect_chain task was used for the above five energy bands with likelihood threshold =8, and energy conversion factors in units $10^{11}$ count. $cm^2$ $erg^{-1}$ . In order to utilize the $\chi^2$ technique, the X-ray spectra were rebinned to contain at least 20 counts in each spectral bin using grppha command and then simultaneously fitting the spectra from MOS1, MOS2 and PN detectors, with the response functions for each detector, using the XSPEC spectral fitting package. The value of the galactic absorption ($N_H$) was found to be $3 \times 10^{20}$ $cm^{-2}$ obtained from FTOOLS NH task. For the search of discrete X-ray sources, the detection metatask edetect_chain is applied to the three EPIC cameras. The above five energy bands were used.

## 3- Spectral Analysis

To identify the X-ray sources detected in MKN 205 field, a search program was carried out to compare positions of objects in our field with x-ray source positions of objects in several catalogues and archives (eg. SIMBAD, NED, USNO, APM-North, .etc.). To improve the reliability of the identifications, the optical magnitude (B and V) are used to calculate the ratio of X-ray flux ($f_x$) to optical flux($f_{op}$), which is given by the following relation (Maccacaro et al. 1988) :

$$\log(\frac{f_x}{f_{op}}) = \log(f_x) + 0.4 m_{op} + 5.37.$$

where mop is the optical magnitude. The optical magnitudes are given from SIMBAD, USNO and APM-North catalogues.

A spectral analysis for three bright non-target sources in MKN 205 field are performed. These sources are classified as, early-type galaxies(at right ascension & declination 12.338&75.370 (named Obj_12.338_75.370)), broad-line active galactic nuclei (BLAGN)( at right ascension & declination 12.348&75.833 (named Obj_12.348_75.833)) and narrow emission line galaxies (NELG)( at right ascension & declination 12.368&75.438 (named Obj_12.368_75.438)).
In this spectral analysis, three sets of data points and model curves (one for PN and two for MOS) were used. For all in the following plots, the upper curves are for PN data and the lower curves are for the MOSs data.

For the object Obj_12.338_75.370 an X-ray spectrum with a single-temperature thermal plasma model (Raymond and Smith 1977) modified by interstellar absorption was fitted to all the range from 0.2 keV to 10.0keV. The parameters : plasma temperature, the metal abundance and the normalization were free in the fitting, and the absorption column density fixed at $3 \times 10^{20}$ cm$^{-2}$.

This model failed to reproduce the observed spectrum, it fitted the data well up to 2 keV. The $\chi^2$/odf is .345.5/211, KT is 0.72±0.015 Kev and the abundance parameter is 0.16±0.02. By using this model an excess emission found above the 2 keV (Fig1). A thermal bremsstrahlung model for the hard component to estimate the temperature variation above 2 keV was assumed. The bremsstrahlung model together with Raymond model was then fitted to the same band (0.2keV - 10.0 keV). The assumption of thermal bremsstrahlung model to this spectrum provided an acceptable fit ( Fig 2) the $\chi^2$/odf is 217/209,the reduced chi-square is 1.042 and the null hypothesis probability is 0.323. The output thermal temperature from this fit is KT 0.41±0.01 Kev, while the thermal temperature from the bremsstrahlung is 5.39±1.6 Kev. In this fit the abundance parameter is 0.21±0.03.

The broad band spectrum in when fitted by Power-law with Raymond model an acceptable fit was obtained, the $\chi^2$/odf = 220/209 with reduced chi-square is 1.052 and the null hypothesis probability is 0.29, with photon index 1.82±0.19. The thermal temperature from Raymond- Power-law model is 0.40±0.01 as from Raymond- Bremsstrahlung model (Fig 3). Therefore, we cannot distinguish between the two possible thermal and non-thermal models of the hard component from the spectrum alone.

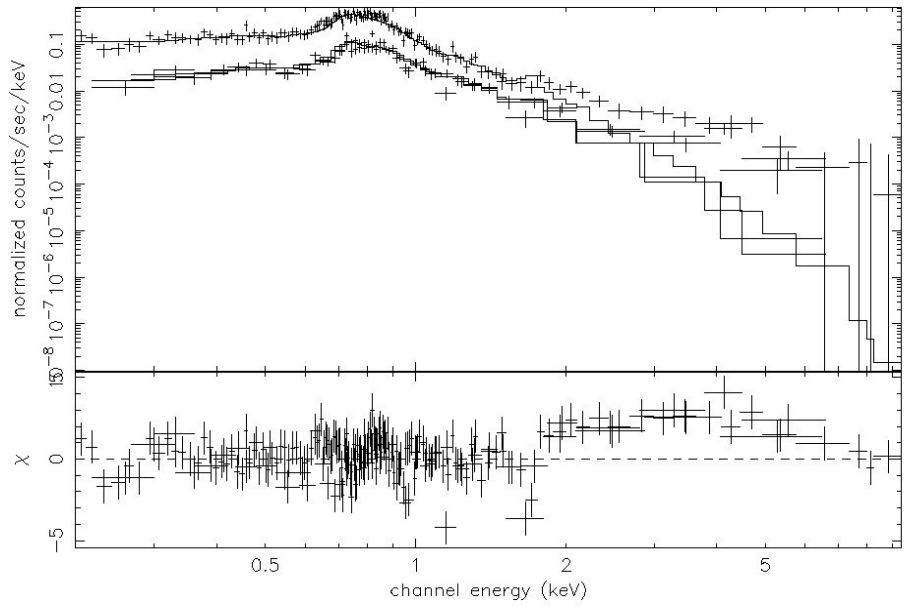

Figure 1: The PN, MOS1 and MOS2 spectra fitted using Raymond thermal model, with $\chi^2$ Test.

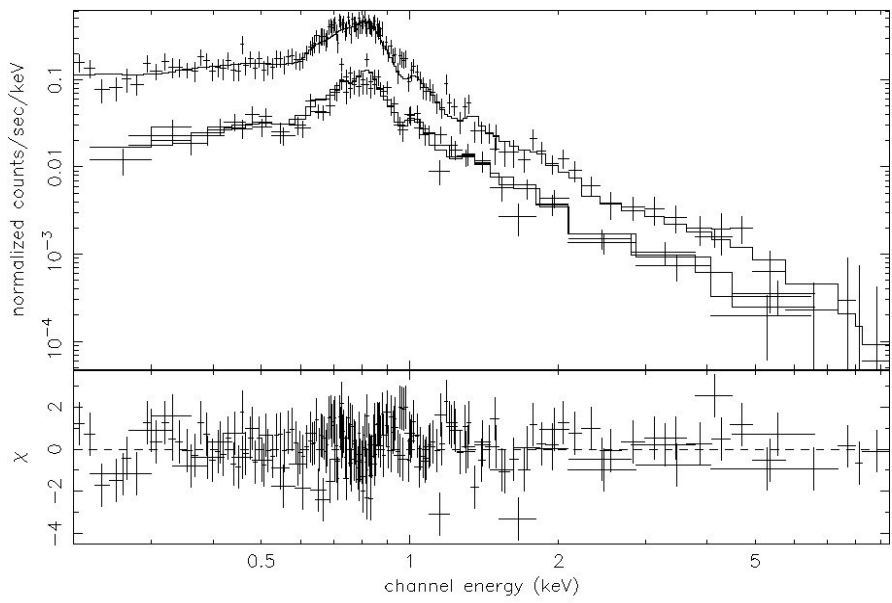

Figure 2: The PN, MOS1 and MOS2 spectra fitted using Raymond thermal model with thermal Bremsstrahlung, with $\chi^2$ Test.

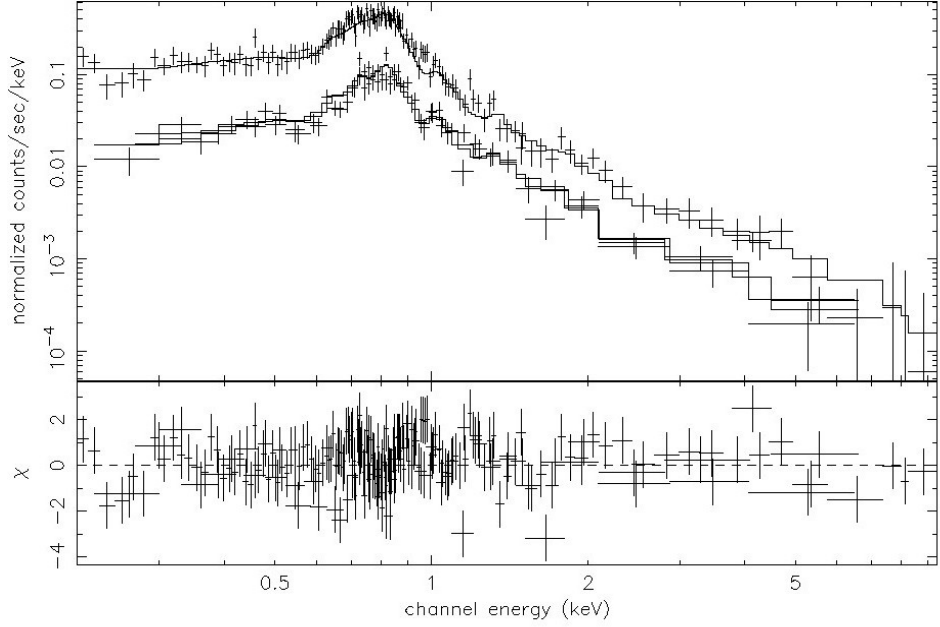

Figure 3: The PN, MOS1 and MOS2 spectra fitted using Raymond thermal model and Power-law model, with $\chi^2$ Test.

The two other objects Obj_12.348_75.833 and Obj_12.368_75.438 are classified as BLAGN and NELG respectively. The spectral analysis of these objects started with fitting a single power-law to the background-subtracted spectra. This model has two free parameters, the normalisation, and the continuum slope $\Gamma$. A fixed photoelectric absorption component was included to account for the effect of the galactic absorption along the line of sight. Fitting this model gives a good reduced $\chi^2$ for the two objects(Fig 4 and 5) with photon index 1.7±0.04 and 1.5±0.03 respectively. The fitted parameters are summaried in tables (2).

Table 2 the fit parameters of the power-law model for the two objects

| Parameter | Obj_12.348_75.833 | Obj_12.368_75.438 |
|---|---|---|
| Absorption column density ($N_H$) | $3 \times 10^{20}$ cm$^{-2}$ (faxed) | $3 \times 10^{20}$ cm$^{-2}$ (faxed) |
| $\Gamma$ | 1.7±0.04 | 1.5±0.03 |
| $\chi^2$ | 87 | 153 |
| odf | 99 | 138 |
| Norm | $1\times10^{-4}$ | $8.3\times10^{-5}$ |

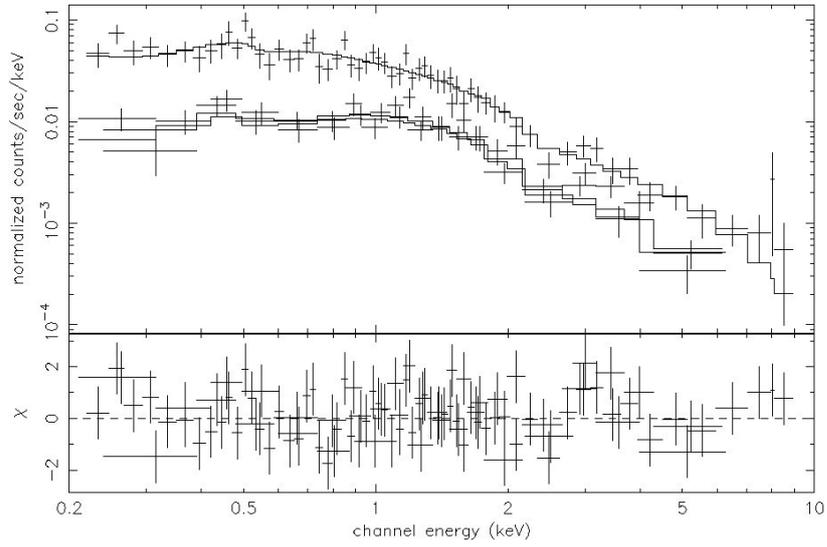

Figure 4: The PN, MOS1 and MOS2 spectra fitted using Power-law model with fixed photoelectric absorption component for Obj_12.348_75.833

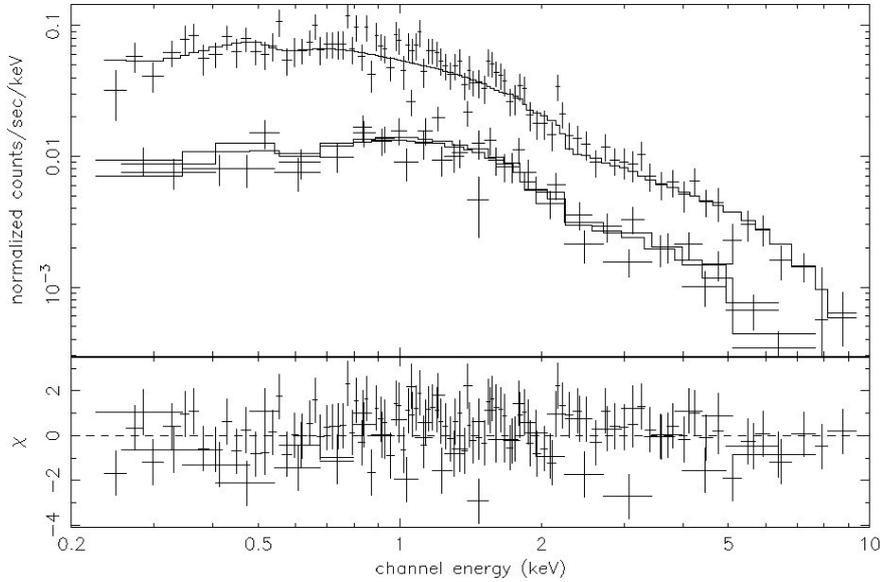

Figure 5: The PN, MOS1 and MOS2 spectra fitted using Power-law model with fixed photoelectric absorption component for Obj_12.368_75.438

In order to test whether intrinsic absorption is present, photoelectric absorption at the redshift of the source (0.65 and 0.24) is fitted where, the redshift was taken from Barcons, et al. (2002). In this case three parameters are free, the normalisation, the photon index $\Gamma$, and the rest frame absorption. There is no significant improvement by adding this component, in the reduced $\chi^2$(0.89 and 1.07 respectively). The intrinsic absorption components(NH) from this fit are $1.13\times10^{11}$ ±0.05 and $3.53\times10^{11}$ ±0.015, it does not make sense. It is not possible to measure NH values much lower than $10^{20}$ cm$^{-2}$ with XMM. This means that we need data with much better statistics to simultaneously fit the different contributions of intrinsic absorption for this object. The F-test ( F = 5.8, F-probability

= 1.73 ×$10^{-2}$ ) tells us that going from power-law model with fixed NH to a power-law model with intrinsic NH as additional fit parameter represents a significant improvement of the fit (Fig 6-7). The agreement for this object is excellent so, one can assume that the flux distribution is affected by intrinsic absorption.

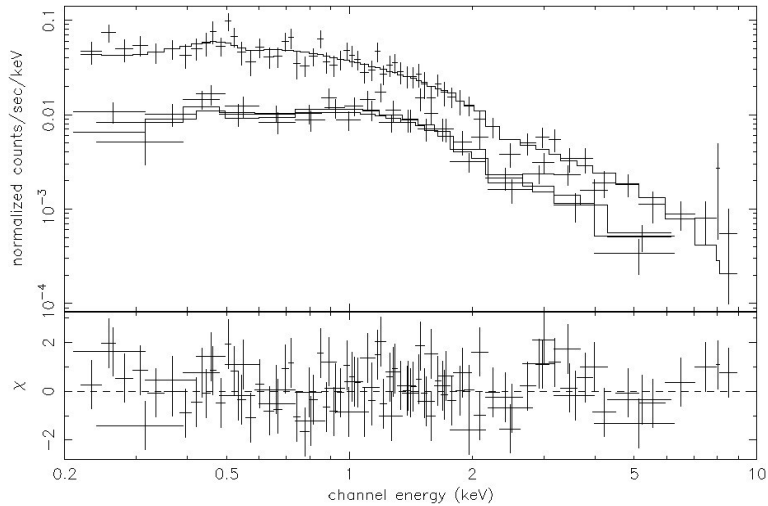

Figure 6: The PN, MOS1 and MOS2 spectra fitted using Power-law model with, photoelectric absorption at the redshift for Obj_12.348_75.833

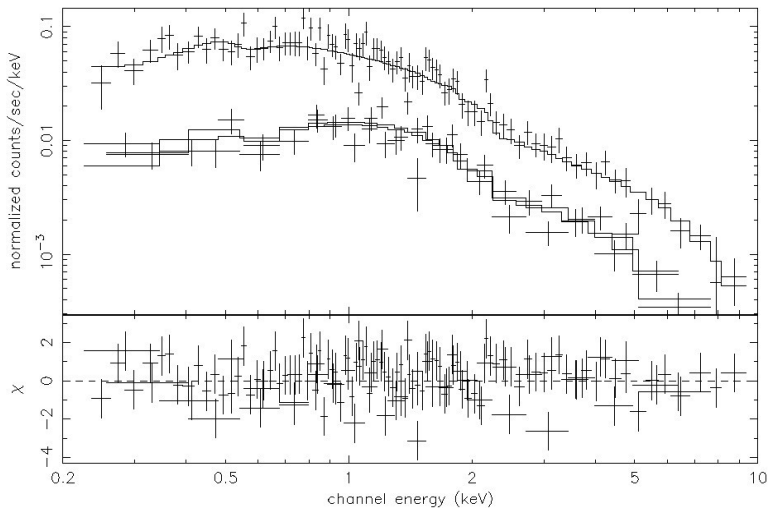

Figure 7: The PN, MOS1 and MOS2 spectra fitted using Power-law model with photoelectric absorption at the redshift for Obj_12.368_75.438

The determination of soft excess may depend on the knowledge of the shape of the power-law and the quantity of absorption. It has been interpreted as a primary emission from the accretion disc, the gravitation energy released by emission from the accretion disc, the gravitation energy released by viscosity in the disc or as secondary radiation from the reprocessing of hard X-rays in the surface layers of the disc. We can provide a good fit to this soft excess by several models, such as single black bodies, multiple black bodies, multicolor dick black body, bluried reaction from partially ionized material, smeared absorption, and thermal computerization in the optically thick medium (B. Korany and M. Nouh 2019). To do this, the spectra were fitted with power-law and a low energy black body component (at the redshift of the source) taking into account absorption in our Galaxy. The $\chi^2$/odf are 85.05/97 and 149/136 respectively for these fits. The photon indexes are changed to 1.42±0.04 and 1.42±0.05 and the thermal temperatures from the black body component (KT) are 0.39±0.17 and 0.38±0.17 Kev (Table 3 summaries the fit parameters) as shown in figures (8 and 9).

The F-test (F 1.01, 1.90 and F-probability 0.368, 0.153) tells us that going from power-law model to a power-law model with black body component as an additional fit parameter does not represent a significant improvement of the fit. This means that we need more data with much better statistics to simultaneously fit the different contributions of soft excess component.

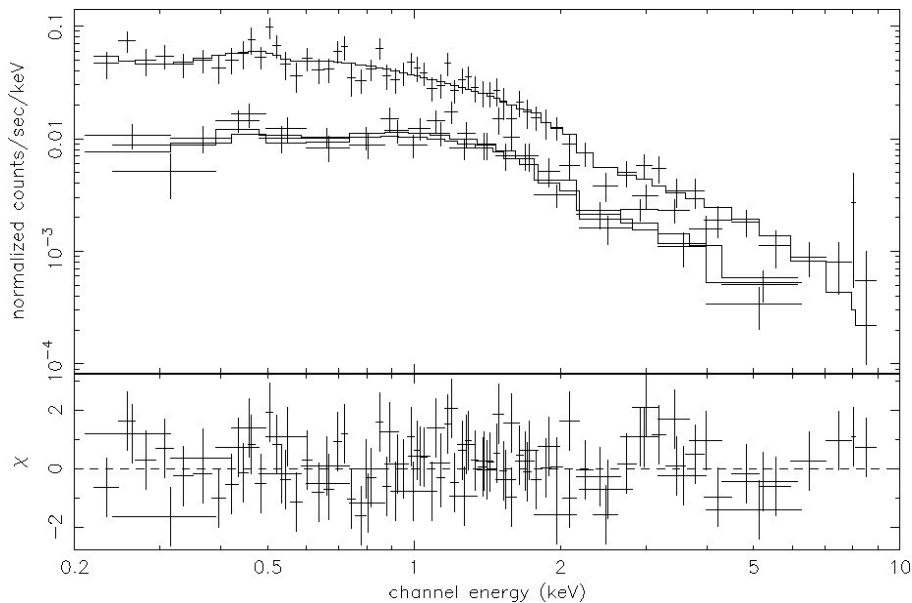

Figure 8: The PN, MOS1 and MOS2 spectra fitted using Power-law model with a black-body component for Obj_12.348_75.833.

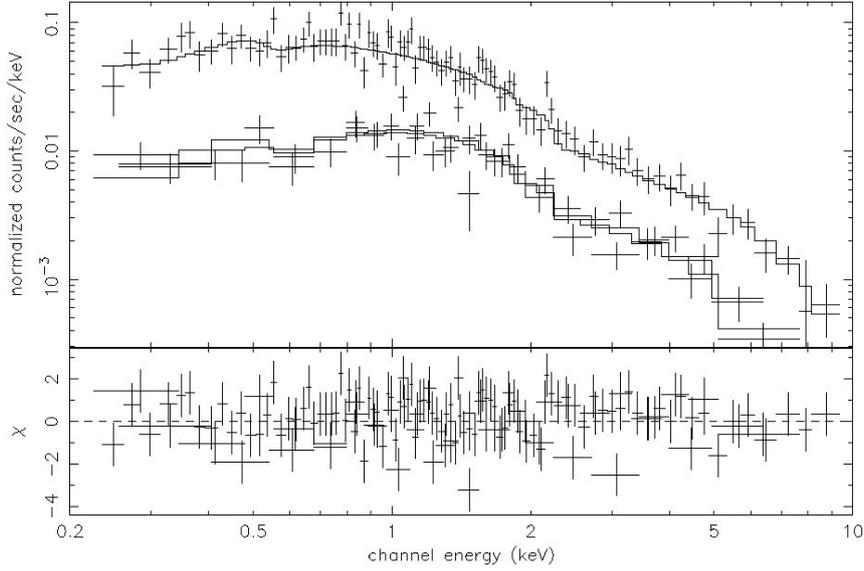

Figure 7: The PN, MOS1 and MOS2 spectra fitted using Power-law model with a black-body component for Obj_12.368_75.438

Table 3 the fit parameters of the power-law model with a black-body component for the two objects

| Parameter | Obj_12.348_75.833 | Obj_12.368_75.438 |
|---|---|---|
| Absorption column density ($N_H$) | $3 \times 10^{20}$ cm$^{-2}$ (faxed) | $3 \times 10^{20}$ cm$^{-2}$ (faxed) |
| Γ | 1.6±0.06 | 1.4±0.05 |
| KT | 0.39±0.17 Kev | 0.38±0.17 |
| $\chi^2$ | 85 | 149 |
| odf | 97 | 136 |
| Norm | 1.45×10$^{-7}$ | 4.68×10$^{-7}$ |

4- **Discussion and Conclusion**

The aim of the paper was to detect, classify and spectral analyses the possible bright X-ray non-target sources in MKN 205 field. We detected three bright objects in this field, these sources are located at right ascension & declination 12.338&75.370, 12.348&75.833 and 12.368&75.438 and classified optically as Early-type galaxy, BLAGN, and NELG respectively.

The flaring high background periods checked by creating alight curves. For the first object an X-ray spectrum with a single-temperature thermal plasma model modified by interstellar absorption was fitted to from 0.2 keV to 10.0keV with fixed the absorption column density at $3 \times 10^{20}$ cm$^{-2}$. The thermal temperature ( KT ) of this fit is 0.72±0.015 Kev and the abundance parameter are 0.16±0.02. A thermal bremsstrahlung model for the hard component to estimate the temperature variation above 2 keV was assumed. The bremsstrahlung model together with Raymond model was then fitted to the same band. The assumption of thermal bremsstrahlung model to this spectrum provided an acceptable fit and the output thermal temperature from this fit is 0.41±0.01 Kev, while the thermal temperature from the bremsstrahlung is 5.39±1.6 Kev. In this fit the abundance parameter is 0.21±0.03. When the spectrum fitted by Power-law with Raymond model, an acceptable fit was obtained with a photon index 1.82±0.19 and the thermal temperature from Raymond- Power-law model is 0.40±0.01 as from Raymond- Bremsstrahlung model.

The two other objects fitted firstly by a single power-law to the background-subtracted spectra with a fixed photoelectric absorption component was included to account for the effect of the galactic absorption along the line of sight. A photoelectric absorption at the redshift of the source fitted in order to test whether intrinsic absorption is present. There is no significant improvement by adding this component. The intrinsic absorption components (NH) from this fit are $1.13 \times 10^{11}$ ±0.05 and $3.53 \times 10^{11}$ ±0.015. It is not possible to measure NH values much lower than $10^{20}$ cm$^{-2}$ with XMM. The F-test tells us that going from power-law model with fixed NH to a power-law model with intrinsic NH as an additional fit parameter represents a significant improvement of the fit. The agreement for this object is excellent so, one can assume that the flux distribution is affected by intrinsic absorption. Finally, we tested the X-Ray soft excess by adding a black body component to the power law model and the thermal temperatures from the black body component (KT) are 0.39±0.17 and 0.38±0.17 Kev. But The F-test tells us that going from power-law model to a power-law model with black body component as an additional fit parameter does not represent a significant improvement of the fit.